\documentclass{PoS}
\usepackage{graphicx}

\title{Computing the long-distance contribution to 
second order weak amplitudes}

\ShortTitle{Long-distance contributions to weak amplitudes}

\author{\speaker{Norman H. Christ}\\
        Department of Physics, Columbia University, New York, NY 10025, USA \\
        E-mail: \email{nhc@phys.columbia.edu}}

\author{RBC and UKQCD collaborations}

\abstract{
The calculation of the long-distance contribution to the 
$K^0-\overline{K}^0$ mass matrix is divided into three parts:  
First, the calculation of the matrix element between kaon 
states of the product of two space-time integrated, 
$\Delta S=1$, four-quark weak operators.  Second an RI/MOM 
subtraction to remove the short distance part of this matrix 
element in a fashion consistent with the calculation of the 
physical short distance part.  Third an application of the 
Lellouch-Luscher method, generalized to second order in the 
weak interactions, to control finite volume errors.  Such 
an approach promises to permit accurate lattice calculation 
of the $K_L$-$K_S$ mass difference and the long-distance 
contributions to $\epsilon_K$.}

\FullConference{The XXVIII International Symposium on Lattice Field Theory, Lattice2010\\
		June 14-19, 2010\\
		Villasimius, Italy}

\begin{document}

\section{Introduction}

Lattice QCD has been very successful at computing the effects of the
electroweak interactions on the properties of the strongly interacting
particles.  For many processes the large mass of the $W^\pm$ and $Z$
bosons cause their interactions with the quarks and gluons of the hadrons
to take place in a very small space-time region.  These short distance 
interactions can be evaluated using electroweak and QCD perturbation theory 
and their low energy effects on hadrons described by effective four quark 
operators.  For example, this approach provides a good description of 
both first-order decays and even some second order processes such as the 
CP violating effects in $K^0 - \overline{K}^0$ and $B^0 - \overline{B}^0$ 
mixing.  However, for general second order processes, in which two $W^\pm$ 
and/or $Z$ bosons appear, it is possible that while each $W^\pm$ or $Z$ exchange 
will appear to take place at a point, the points locating these two exchanges 
may be separated by a much larger distance $\sim 1/\Lambda_{\rm QCD}$.  
Such long distance effects are believed to contribute to the CP violation 
seen in $K^0 - \overline{K}^0$ mixing at the 5\% level~\cite{Buras:2010pza} 
but on at least the 20\% level~\cite{Herrlich:1993yv} for the CP 
conserving $K_L - K_S$ mass difference.\footnote{For a discussion of the
lattice QCD calculation of long distance effects in different decay 
processes see Ref.~\cite{Isidori:2005tv}.}

Here we present a method to compute such long distance effects using
lattice QCD, focused on the case of the $K_L - K_S$ mass difference.  
There are three complications which must be overcome.  First we need to 
devise an Euclidean space expectation value which can be evaluated in
lattice QCD and which contains the second order energy shift of interest.

Second, such a lattice quantity will involve a product of two, first-order
weak Hamiltonian densities, ${\cal H}_W(x_i)_{i=1,2}$, each corresponding 
to one of the $W^\pm$ or $Z$ exchanges.   The short distance behavior of
this product as $|x_1-x_2| \rightarrow 0$ will not describe the actual
behavior of the exchange of two $W^\pm$ or $Z$ bosons at nearby 
space-time points.  Thus, this incorrect short distance behavior must
be removed and replaced by the known, physical, short distance behavior 
described above.

Third, the effects of finite volume, necessary in a lattice calculation,
must be removed.  These appear especially significant since the infinite
volume expression contains continuous integrals, often with vanishing
energy denominators evaluated as principal parts, while the finite volume
quantity is a simple sum of discrete finite volume states.  Here a
generalization of the method of Lellouch and Luscher~\cite{Lellouch:2000pv} 
can be used.  We will now discuss how each of these obstacles may be 
overcome in a calculation of the $K_L - K_S$ mass difference, $\Delta m_K$.

\section{Second order lattice amplitude}
The standard description of $K^0 - \overline{K}^0$ mixing provides an 
expression for the $K_L - K_S$ mass difference which we will write as
\begin{equation}
\Delta m_K = 2{\cal P} \sum_\alpha \frac{\langle \overline{K}^0 | H_W|\alpha\rangle 
                                     \langle \alpha|H_W|K^0\rangle}
                                        {m_K - E_\alpha}.
\label{eq:delta_m_IV}
\end{equation}
Here CP violating effects, at the 0.1\% level, have been neglected, we are 
summing over intermediate states $|\alpha\rangle$ with energy $E_\alpha$ 
and normalization factors associated with the conserved total momentum are 
suppressed.  This generalized sum includes an integral over intermediate 
state energies and the $\cal P$ indicates the principal part of the integral 
over the $E_\alpha=m_K$ singularity.

\begin{figure}[ht]
\centering
\includegraphics[width=0.7\textwidth]{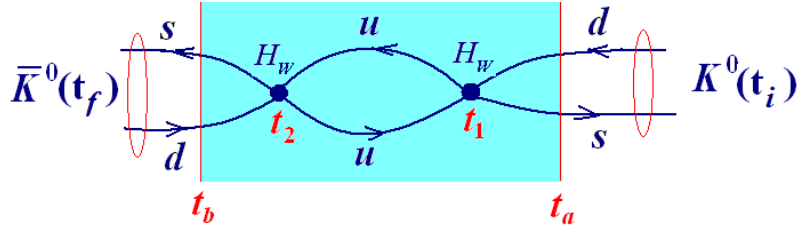}
\vskip -0.1in
\caption{One type of diagram contributing to $\cal A$ of
Eq.~\protect\ref{eq:lattice_amplitude}.  Here $t_2$ and $t_1$ 
are integrated over the interval $[t_a,t_b]$, represented by the shaded 
region between the two vertical lines.  In addition to this connected 
quark flow there will also be disconnected diagrams in which no quark 
lines connect $H_W(t_2)$ and $H_W(t_1)$.}
\label{fig:lattice_amplitude}
\end{figure}

One possible way to capture a similar expression in a Euclidean space
lattice calculation is to evaluate the time-integrated second-order product
that if evaluated in Minkowski space would yield the $\Delta m_K$ 
contribution to the time evolution over a time interval $[t_a,t_b]$:
\begin{equation}
{\cal A} = \frac{1}{2}\langle \overline{K}^0(t_f) \int_{t_a}^{t_b} d t_2 
                  \int_{t_a}^{t_b} d t_1
                          H_W(t_2) H_W(t_1) K^0(t_i) \rangle.
\label{eq:lattice_amplitude}
\end{equation}
Here the initial $K^0$ is created by a source $\overline{K}^0(t_i)$ at the 
time $t_i$ and the final $\overline{K}^0$ state destroyed by the sink 
$\overline{K}^0(t_f)$ at time $t_f$.  This amplitude is represented 
schematically in Fig.~\ref{fig:lattice_amplitude}. 
Equation~\ref{eq:lattice_amplitude} can be evaluated as a standard Euclidean 
space path integral with $t_f \gg t_b \gg t_a \gg t_f$.  If the time extent 
of this Euclidean path integral is sufficiently large, then when converted 
to an operator expression, Eq.~\ref{eq:lattice_amplitude} becomes the vacuum 
expectation value of the time-ordered product of Heisenberg operators.  
Assuming that $t_f-t_b$ and $t_a-t_i$ are sufficiently large to project 
onto the $\overline{K}^0$ and $K^0$ states, substituting a sum over energy 
eigenstates $|n\rangle$, and integrating over $t_2$ and $t_1$ one obtains:
\begin{eqnarray}
{\cal A} &=& -\sum_{n \ne n_0} \frac{\langle \overline{K}^0 |H_W|n\rangle 
                                     \langle n|H_W|K^0\rangle}
                                        {m_K - E_n}
          \left\{t_b-t_a - \frac{e^{-(E_n-m_K)(t_b-t_a)} - 1}{m_K-E_n}\right\}
          e^{-(t_f-t_i)m_K}
\nonumber \\
         &&\quad  -\frac{1}{2}(t_b-t_a)^2\langle\overline{K}^0|H_W|n_0\rangle
                             \langle n_0|H_W| K^0\rangle e^{-(t_f-t_i)m_K}.
\label{eq:lattice_amplitude_explicit}
\end{eqnarray}
Anticipating a result from Sec.~\ref{sec:finite_volume}, we have assumed 
that a single two-pion intermediate state $|n_0\rangle$ is degenerate with 
the kaon and treated that state separately in the time integrations.

The coefficient of the $(t_b-t_a)$ term in Eq.~\ref{eq:lattice_amplitude_explicit} 
is then a finite volume approximation to $\Delta m_K$:
\begin{equation}
\Delta m_K^{\rm FV} = 2\sum_{n \ne n_0} \frac{\langle \overline{K}^0 |H_W|n\rangle 
                                     \langle n|H_W|K^0\rangle}
                                        {m_K - E_n}.
\label{eq:delta_m_FV}
\end{equation}
  The other terms in Eq.~\ref{eq:lattice_amplitude_explicit} fall into 
four categories:  i) The term independent of $t_b-t_a$ within the large 
curly brackets.  This constant must be distinguished from the desired 
term proportional to $t_b-t_a$.  ii) Exponentially decreasing terms 
coming from states $|n\rangle$ with $E_n > m_K$.  These should be negligible 
if $t_b-t_a$ is sufficiently large.  iii)  Exponentially increasing 
terms coming from states $|n\rangle$ with $E_n < m_K$.  These will be 
the dominant contributions and must be accurately determined and 
removed as discussed in the paragraph below\footnote{The author thanks 
Guido Martinelli and Stephen Sharpe for pointing out this behavior 
which had been overlooked when this talk was presented.}.  iv) The 
final term proportional to $(t_b-t_a)^2$ arises because our choice of 
volume makes one $\pi-\pi$ state, $|n_0\rangle$, degenerate with the kaon.

The exponentially growing terms pose a significant challenge.  Fortunately, 
we have some freedom
to reduce their number and complexity.  The two leading terms corresponding
to the vacuum and single pion states can be computed separately and subtracted.
Two pion states lying below $m_K$ can be eliminated using the same 
techniques that have been developed to evade the Maiani-Testa theorem and 
force the lowest energy $\pi-\pi$ state to be the on-shell 
$K \rightarrow \pi\pi$ decay product.  Either choosing
the kaon to have a non-zero laboratory momentum of 753 MeV or introducing
G-parity boundary conditions to force non-zero pion momentum can eliminate 
all $\pi-\pi$ states with energy below $m_K$, at least for those lattice 
volumes that will be accessible within the next few years.

\section{Short distance correction}
\label{sec:short_distance}

The product of operators appearing in Eq.~\ref{eq:lattice_amplitude}
accurately describes the second order weak effects when the corresponding
Hamiltonian densities ${\cal H}(x_i)_{i=1,2}$ are evaluated at space-time 
points separated by a few lattice units $a$: $|x_2 - x_1| \gg a$.
However, as $|x_2 - x_1| \rightarrow 0$ the behavior is unphysical,
being dominated by lattice artifacts rather than revealing the short 
distance structure of $W^\pm$ and $Z$ exchange.  Fortunately, 
non-perturbative Rome-Southampton methods can be applied here to 
accurately remove this incorrect behavior and replace it with the
correct short distance behavior, that portion of the process that
has been traditionally computed using lattice methods.

\begin{figure}[ht]
\centering
\includegraphics[width=0.3\textwidth]{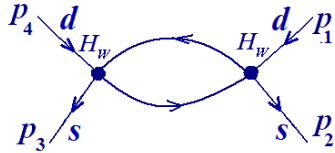}

\caption{Diagram representing a class of connected contributions to 
the off-shell, four-quark Green's function defined in Eq.~\protect\ref{eq:NPR}.
In a non-perturbative evaluation of the Green's function $\Gamma(p_i)$,
graphs of this sort including all possible gluon exchanges would be
included.}
\label{fig:NPR}
\end{figure}

This can be done by identifying the short distance part of the 
amplitude by evaluating the four-quark, off-shell Green's function
\begin{equation}
\Gamma_{\alpha\beta\gamma\delta}(p_i)
    =\langle \widetilde{\overline{d}}_\alpha(p_4) \widetilde{s}_\beta(p_3)
      \int d^4x_1 d^4 x_2 {\cal H}_W(x_2){\cal H}_W(x_1)
      \widetilde{s}_\gamma(p_2)\widetilde{\overline{d}}_\delta(p_1)\rangle.
\label{eq:NPR}
\end{equation}
Here the quark fields are Fourier transformed and the gauge is fixed. 
A class of connected contributions to this Green's function is shown
in Fig.~\ref{fig:NPR}.  A standard application of Weinberg's theorem
demonstrates that if the external momenta $p_i$ obey a condition such
as $p_i \cdot p_j = \mu^2(1-4\delta_{ij})$, then for 
$\mu^2 \gg \Lambda_{\rm QCD}^2$ all of the internal momenta contributing
to $\Gamma(p_i)$ will have the scale $\mu$, up to terms of order
$\Lambda_{\rm QCD}^2/\mu^2$.

At low energies this high momentum part of the integrated product 
${\cal H}_W(x_2){\cal H}_W(x_1)$ can be represented as a linear
combination of four-quark operators $\{O_s\}_{1 \le s \le S}$.  
These operators are typically normalized by imposing conditions on
off-shell Green's functions similar to that in Eq.~\ref{eq:NPR} in 
which the product of ${\cal H}_W$ operators is replaced by $O_s$ and 
the same kinematic point evaluated.  The result is an alternative 
expression for the short distance part of the amplitude $\cal A$:
\begin{equation}
{\cal A}_{\rm SD} = \langle \overline{K}^0(t_f) \int_{t_a}^{t_b} d x_0 \int d^3 x
        \sum_{s=1}^S c_s^{\rm lat}(\mu^2) O_s(\vec x, x_0) K^0(t_i) \rangle.
\label{eq:lattice_amplitude_SD}
\end{equation}
The functions $c_s^{\rm lat}(\mu^2)$ are Wilson coefficients  
for the lattice-regularized operator product expanded in operators 
normalized using the regularization invariant (RI) Rome-Southampton 
scheme.

Thus, we can replace
the incorrect short distance part of our lattice operator product by
the correct continuum contribution by adding to the integrated operator 
product in Eq.~\ref{eq:lattice_amplitude} the operator:
\begin{equation}
\int_{t_a}^{t_b}d x_0 \int d^3 x \sum_{s=1}^S
\left\{c_s^{\rm cont}(\mu^2) - c_s^{\rm lat}(\mu^2)\right\}O_s(\vec x,x_0).
\label{eq:SD_correction}
\end{equation}
Here the $\{c_s^{\rm cont}\}_{1 \le s \le S}$ are the usual continuum 
Wilson coefficients that are computed from electroweak and QCD 
perturbation theory to represent the correct short distance part of the 
physical second order weak process while the lattice coefficients 
$c_s^{\rm lat}$ can be computed from the somewhat elaborate but well 
defined lattice RI/MOM calculation of the Green's functions in 
Eq.~\ref{eq:NPR}.

An important issue on which the above argument depends is the degree
to which the dimension-6, four-quark operators introduced above
capture the entire short distance part of the lattice amplitude.  
Since the degree of divergence of the diagram shown in Fig.~\ref{fig:NPR} 
is +2, the integration that remains after the ``subtraction'' of 
the operator in Eq.~\ref{eq:SD_correction} will still receive $O(1)$
contributions from the lattice scale.  This difficulty can be avoided 
by including dimension eight terms in the Wilson expansion employed in 
Eq.~\ref{eq:SD_correction}.  A more physical and more practical 
approach includes the charm quark in the lattice calculation so 
that GIM suppression makes the integration more convergent.  

\section{Controlling finite volume errors}
\label{sec:finite_volume}

We now turn to the heart of this proposal: a demonstration that the
potentially large volume dependence coming from those energy denominators
in Eq.~\ref{eq:lattice_amplitude_explicit} with $E_n \sim m_K$ can
be removed, leaving $O(1/L^4)$ finite-volume errors.
This important conclusion is a consequence of a generalization
of the original method of Lellouch and Luscher.  The starting point
is Luscher's relation~\cite{Luscher:1990ux} between an allowed, 
finite-volume, two-particle energy, $E = 2\sqrt{k^2 +  m_\pi^2}$ and 
the two-particle scattering phase shift $\delta(E)$:  
\begin{equation}
\phi(kL/2\pi) +\delta(E) = n\pi
\label{eq:Luscher}
\end{equation}
where $n$ is an integer and the known function $\phi(q)$ is defined 
in Ref.~\cite{Luscher:1990ux}.

Following Lellouch and Luscher we consider the $s$-wave, $\pi-\pi$ 
scattering phase shift as modified by the weak interactions and 
use Eq.~\ref{eq:Luscher} to connect this to the finite volume energies, 
determined using degenerate perturbation theory for the 
$K_S \leftrightarrow \pi-\pi$ finite volume system.  For simplicity 
we will limit our discussion to the larger $\Delta I = 1/2$ part of 
$H_W$ and the $I=0$ $\pi-\pi$ state.\footnote{Treating the general 
case is not difficult: $H_W^{\Delta I=3/2}\cdot H_W^{\Delta I=3/2}$ 
and $H_W^{\Delta I=1/2}\cdot H_W^{\Delta I=1/2}$ can be analyzed in 
the same way while the combination $H_W^{\Delta I=1/2}\cdot H_W^{\Delta I=3/2}$ 
contains no two-pion intermediate states.}  The relation between the
finite and infinite volume second order mass shift is obtained by 
imposing Eq.~\ref{eq:Luscher}, accurate through second order in $H_W$.

We begin by examining the energies, accurate through second order in
the strangeness changing, $\Delta I = 1/2$ weak Hamiltonian, $H_W$, 
of the finite volume system made up of a $K_S$ meson, an $I=0$ 
two-pion state $|n_0\rangle$ with energy $E_{n_0}$ nearly degenerate 
with $m_K$ and other single and multi-particle states coupled to 
$K_S$  and $|n_0\rangle$ by $H_W$.  Following second order degenerate 
perturbation theory, we can obtain the energies of the $K_S$ and 
two-pion state $|n_0\rangle$ as the eigenvalues of the $2 \times 2$ 
matrix:
\begin{equation}
\left( \begin{array}{cc}
 m_K + \sum_{n \ne n_0}
               \frac{|\langle n|H_W|K_S\rangle|^2}{m_K -E_n}
       & \langle K_S|H_W|n_0\rangle \\
\langle n_0 |H_W | K_S\rangle & E_{n_0}+\sum_{n \ne K_S}
               \frac{|\langle n|H_W|n_0\rangle|^2}{E_{n_0} -E_n}
\end{array}\right).
\label{eq:fv_2x2}
\end{equation}

\begin{figure}[ht]
\centering
\includegraphics[width=1.0\textwidth]{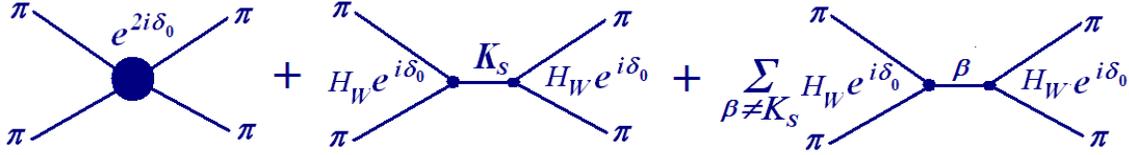}
\caption{Diagrams showing the three contributions to the $\pi-\pi$
phase shift when both strong and second order weak effects are 
included.  The states $\beta$ are multi-particle states with $S=\pm 1$.}
\label{fig:resonance}
\end{figure}

Finite and infinite volume quantities can then be related by requiring 
that the eigenvalues of the $2 \times 2$ matrix in Eq.~\ref{eq:fv_2x2} 
solve Eq.~\ref{eq:Luscher} where the phase shift $\delta(E)$ is the 
sum of that arising from the strong interactions, $\delta_{\,0}(E)$, a 
resonant contribution from the $K_S$ pole and more familiar 
second-order Born terms:
\begin{equation}
\delta(E) = \delta_{\,0}(E) + \arctan(\frac{\Gamma(E)/2}{m_K+\Delta m_{K_S}-E}) 
       -\pi\sum_{\beta \ne K_S} \frac{|\langle \beta|H_W|n_0\rangle|^2}{E -E_\beta}.
\label{eq:delta_second_order}
\end{equation}
Here $\Gamma(E)$ is proportional to the square of the $K_S$ - two pion 
vertex which becomes the $K_S$ width when evaluated at $E=m_K$:
\begin{equation}
\Gamma(E) = 2\pi |\langle \pi\pi(E)|H_W|K_S\rangle|^2,
\end{equation}
where for the infinite volume, $I=0$, $s$-wave, 2-pion state we choose 
the convenient normalization $\langle \pi\pi(E)|\pi\pi(E')\rangle = \delta(E-E')$.  
The three terms in Eq.~\ref{eq:delta_second_order} are shown in 
Fig.~\ref{fig:resonance}.

The easiest case to examine is that in which the volume is chosen
to make $E_{n_0}-m_K$ very small on the scale of $\Lambda_{\rm QCD}$
but large compared to $\Gamma$ or $\Delta m_K$, so that $m_K$ and
$E_{n_0}$ are not ``degenerate''.  Expanding Eq.~\ref{eq:Luscher} and the 
$\pi-\pi$ energy eigenvalue from Eq.~\ref{eq:fv_2x2} in $H_W$ and 
collecting all terms of second order in $H_W$ we find:
\begin{eqnarray}\mbox{\ }\hskip -0.2in
\left.\frac{\partial \Bigl(\phi+\delta_{\,0}\Bigr)}{\partial E}\right|_{E=E_{n_0}}
\hskip -0.1in
\left\{\frac{|\langle K_S|H_W|n_0\rangle|^2}{E_{n_0}-m_K} + \sum_{n \ne K_S}
   \frac{|\langle n|H_W|n_0\rangle|^2}{E_{n_0} -E_n} \right\}
= \frac{\Gamma(E_{n_0})/2}{E_{n_0}-m_K} 
+ \hskip -0.1in \sum_{\beta \ne K_S} 
       \frac{\pi|\langle \beta|H_W|\pi\pi\rangle|^2}{E_{n_0} -E_\beta}.
\label{eq:non-degenerate}
\end{eqnarray}

This relation has two useful consequences.  First we can equate the residues 
of the kaon poles, $E_{n_0}=m_K$ on the left- and right-hand sides.  This 
gives us the original Lellouch-Luscher relation.  Second we can subtract 
the pole terms and equate the remaining parts of Eq.~\ref{eq:non-degenerate} 
evaluated at $E_{n_0}=m_K$.  This second result will be used below to
remove the second-order Born terms.

Finally, closer to the original spirit of Lellouch and Luscher, we substitute
the phase shift $\delta(E)$ from Eq.~\ref{eq:delta_second_order} into 
Eq.~\ref{eq:Luscher} and require that the resulting equation be valid
at the energy eigenvalues $E_\pm$ of the $2\times 2$
matrix in Eq.~\ref{eq:fv_2x2} for a box chosen to make $E_{n_0}=m_K$.
To zeroth order in $H_W$, this relation is the usual Luscher relation 
between $\delta_{\,0}(E)$ and the allowed, finite volume, $\pi-\pi$ 
energy.  When Eq.~\ref{eq:Luscher} is expanded to first order, we reproduce
the standard derivation of Lellouch and Luscher's relation.  Expanding to 
second order in $H_W$ yields the desired relation between the finite and 
infinite volume expressions for the second order weak contribution to the
$K_S$ difference:
\begin{eqnarray}
\Delta m_{K_S}  &=& 
\sum_{n \ne n_0}\frac{|\langle n|H_W|K_S\rangle|^2}{m_K-E_n}
 +\frac{1}{\frac{\partial (\phi+\delta_{\,0})}{\partial E}}
  \Bigg[\frac{1}{2}\frac{\partial^2 (\phi+\delta_{\,0})}{\partial E^2} 
      |\langle n_0|H_W|K_S\rangle|^2 \nonumber \\
&& -\frac{\partial}{\partial E_{n_0}}\left\{
    \left.\frac{\partial(\phi+\delta_{\,0})}{\partial E}\right|_{E=E_{n_0}}\hskip -0.1in 
    |\langle n|H_W|K_S\rangle|^2\right\}\Bigg] 
\label{eq:result}
\end{eqnarray}
where Eq.~\ref{eq:non-degenerate}, evaluated at $E_{n_0}=m_K$ with the pole 
terms subtracted has been used to eliminate the second-order Born terms.
Note the $\partial/\partial E_{n_0}$ appearing in the final term in 
Eq.~\ref{eq:result} must be evaluated by varying the spatial volume which 
determines $E_{n_0}$.\footnote{In 
a one-dimensional example, these derivative terms come naturally from a 
generalization of the usual contour integration relation between finite 
volume sums and infinite volume momentum integrals which includes the 
effects of a double pole arising from the vanishing on-shell energy 
denominator.}

To obtain the $K_L-K_S$ mass difference we first 
observe that the $K_L$ second order mass shift is given by a formula similar 
to Eq.~\ref{eq:result} in which $K_L$ replaces $K_S$ and all but the first
term on the right-hand side is omitted since $K_L$ does not couple to two
pions, assuming CP conservation.  Second if this new equation is 
subtracted from Eq.~\ref{eq:result} the result is similar to Eq.~\ref{eq:result} 
with $\Delta m = \Delta m_{K_S} - \Delta m_{K_L}$
on the left-hand side, the first term on the right-hand side
is simply $\Delta m_K^{FV}$ of Eq.~\ref{eq:delta_m_FV} and the remaining two
$O(1/L^3)$ correction terms on the right hand side of Eq.~\ref{eq:result} 
are unchanged.

\section{Conclusion}

We have proposed a lattice method to compute the $K_L-K_S$ mass difference
in which all errors can be controlled at the percent level.  Both short 
and long distance effects are represented, including a possibly 
$\Delta I = 1/2$-enhanced contribution from $I=0$ two pion states.  Given 
the complexity of the analysis, the importance of physical kinematics 
and the difficulty of the disconnected diagrams this calculation is not 
practical today but may be possible in the next few years.

The author thanks his RBC/UKQCD collaborators for important contributions 
to this work and Laurent Lellouch, Guido Martinelli and Stephen Sharpe
for very helpful discussions.  This work was supported in part by U.S. 
DOE grant DE-FG02-92ER40699.

\bibliography{LD.bib}
\bibliographystyle{JHEP}

\end{document}